\documentclass[acmtog,nonacm]{acmart}
\usepackage{colortbl}
\usepackage{multirow}
\usepackage{varwidth} 
\usepackage{pifont} 
\usepackage{boldline}
\usepackage{amsmath}
\usepackage{xcolor}
\usepackage[ruled]{algorithm2e} 
\usepackage{wrapfig}
\usepackage{bm}
\usepackage{makecell}
\usepackage{arydshln}
\usepackage[utf8x]{inputenc}
\usepackage{siunitx}
\usepackage{afterpage}
\usepackage{graphicx}  
\usepackage[shortcuts]{extdash}
\usepackage{tabularx}
\newcolumntype{P}[1]{>{\centering\arraybackslash}p{#1}}
\AtBeginDocument{%
  }

\begin{document}
\newcommand{\rref}{\mathbf{r}_{\text{0}}}
\newcommand{\xref}{{x}_{\text{0}}}
\newcommand{\yref}{{y}_{\text{0}}}
\newcommand{\zref}{{z}_{\text{0}}}

\newcommand{\rlocal}{\mathbf{r}_{l}}
\newcommand{\xlocal}{{x}_{l}}
\newcommand{\ylocal}{{y}_{l}}
\newcommand{\zlocal}{{z}_{l}}

\newcommand{\cov}{\mathbf{\Sigma}}
\newcommand{\rot}{\mathbf{R}}
\newcommand{\rotc}{\mathbf{C}}
\newcommand{\sca}{\mathbf{S}}
\newcommand{\kvec}{\mathbf{k}}

\definecolor{mg}{rgb}{0.639,0.984,0.722}
\definecolor{my}{rgb}{0.996,0.875,0.643}
\definecolor{mr}{rgb}{0.941,0.561,0.620}
\newcommand{\ccg}{\cellcolor{mg}}  
\newcommand{\ccy}{\cellcolor{my}}  
\newcommand{\ccr}{\cellcolor{mr}}  

\newcommand{\cmark}{\ding{51}}%
\newcommand{\xmark}{\ding{55}}%
\newcommand{\bluecheck}{{\color{blue}\checkmark}}
\newcommand{\blackcheck}{{\color{black}\cmark}}
\newcommand{\redx}{{\Large\color{\red}\xmark} }

\newcommand{\headfontsize}{\scriptsize}
\newcommand{\tablerefsize}{\tiny}
\newcommand{\vertice}{\mathbf{v}}
\newcommand{\normal}{\mathbf{n}}

\newcommand*\colourcheck[1]{%
  \expandafter\newcommand\csname #1check\endcsname{\textcolor{#1}{\ding{51}}}%
}
\colourcheck{red}
\colourcheck{green}

\newcommand*\colourxx[1]{%
  \expandafter\newcommand\csname #1xx\endcsname{\textcolor{#1}{\ding{55}}}%
}
\colourxx{red}
\colourxx{green}

\newcommand*\colourtriangle[1]{%
  \expandafter\newcommand\csname #1triangle\endcsname{\textcolor{#1}{$\varDelta$}}%
}
\colourtriangle{yellow}

\newcommand{\greenvcell}{\ccg \cmark}
\newcommand{\redxcell}{\ccr \xmark}
\newcommand{\yellowtrcell}{\ccy $\varDelta$} 
\newcommand{\greenxcell}{\ccg \xmark}
\newcommand{\redvcell}{\ccr \cmark}


\newcommand{\ctcell}{>{\centering\arraybackslash}}
\newcommand{\gw}[1]{\textcolor{purple}{[GW: #1]}}
\newcommand{\sy}[1]{\textcolor{blue}{[SC: #1]}}
\newcommand{\bc}[1]{\textcolor{red}{[BC: #1]}}
\newcommand{\jy}[1]{\textcolor{olive}{[JY: #1]}}
\newcommand{\edit}[1]{\textcolor{red}{#1}}


\newcommand{\citl}{CITL}
\newcommand{\revised}[1]{{#1}}
\newcommand{\editsy}[1]{\textcolor{red}{#1}}
\newcommand{\prop}{\mathcal{P}}
\newcommand{\firstprop}{\prop_1}
\newcommand{\secprop}{\prop_2}
\newcommand{\prophat}{\widehat{g}}
\newcommand{\firstprophat}{\prophat_1}
\newcommand{\secprophat}{\prophat_2}
\newcommand{\target}{a_{target}}
\newcommand{\phaseslm}{\phi}
\newcommand{\efficiency}{\eta}
\newcommand{\fourier}{\mathcal{F}}
\newcommand{\loss}{\mathcal{L}}
\newcommand{\transfer}{\mathcal{H}}
\newcommand{\fx}{f_x}
\newcommand{\fy}{f_y}
\newcommand{\numpixelx}{N_x}
\newcommand{\numpixely}{N_y}
\newcommand{\phasepool}{\mathbb{P}}
\newcommand{\capturedpool}{\mathbb{C}}
\newcommand{\rgbd}{\mathbb{RGBD}}
\newcommand{\lossbp}{\texttt{loss\_bp}}
\newcommand{\fbp}{\texttt{f\_bp}}
\newcommand{\gbp}{\texttt{g\_bp}}
\newcommand{\red}[1]{\textcolor{red}{#1}}
\newcommand{\idx}{i}
\newcommand{\maxidx}{N}
\newcommand{\angularspectrum}{\hat{u}}
\newcommand{\pixelcoord}{\mathbf{r}}
\newcommand{\transmittance}{\mathcal{T}}

\newcommand{\coordinate}{\mathbf{x}}
\newcommand{\freqcoords}{\mathbf{k}}
\newcommand{\mean}{\bm{\mu}}

\newcommand{\worldspace}{w}
\newcommand{\viewspace}{r}
\newcommand{\canonicalspace}{c}
\newcommand{\objectspace}{o}
\newcommand{\centergaussian}{{\bm{\mu}}}
\newcommand{\projectivemapping}{\mathbf{m}}

\newcommand{\mat}[1]{\mathbb{R}^{{#1}\times{#1}}}
\newcommand{\arr}[1]{\mathbb{R}^{{#1} \times 1}} 

\makeatletter
\newcommand\xleftrightarrow[2][]{%
  \ext@arrow 9999{\longleftrightarrowfill@}{#1}{#2}}
\newcommand\longleftrightarrowfill@{%
  \arrowfill@\leftarrow\relbar\rightarrow}
\makeatother

\title{Random-phase Wave Splatting of Translucent Primitives for Computer-generated Holography}

\author{Brian Chao}
\email{brianchc@stanford.edu}
\orcid{0000-0002-4581-6850}
\authornote{denotes equal contribution.}
\affiliation{
  \institution{Stanford University}
  \country{USA}
}
\author{Jacqueline Yang}
\email{jyang01@stanford.edu}
\orcid{0009-0002-3101-3026}
\authornotemark[1]
\affiliation{
  \institution{Stanford University}
  \country{USA}
}
\author{Suyeon Choi}
\email{suyeon@stanford.edu}
\orcid{0000-0001-9030-0960}
\affiliation{
  \institution{Stanford University}
  \country{USA}
}
\author{Manu Gopakumar}
\email{manugopa@stanford.edu}
\orcid{0000-0001-9017-4968}
\affiliation{
  \institution{Stanford University}
  \country{USA}
}
\author{Ryota Koiso}
\email{ryotak@stanford.edu}
\orcid{0000-0001-6160-0366
}
\affiliation{
  \institution{Stanford University}
  \country{USA}
}
\affiliation{
  \institution{KDDI Research}
  \country{Japan}
}
\author{Gordon Wetzstein}
\email{gordon.wetzstein@stanford.edu}
\orcid{0000-0002-9243-6885}
\affiliation{
  \institution{Stanford University}
  \country{USA}
}


\renewcommand{\shortauthors}{Chao, Yang et al.}

\begin{abstract}
Holographic near-eye displays offer ultra-compact form factors for VR/AR systems but rely on advanced computer-generated holography (CGH) algorithms to convert 3D scenes into interference patterns on spatial light modulators (SLMs). However, conventional CGH algorithms typically generate smooth-phase holograms, limiting their ability to capture view-dependent effects and realistic defocus blur while severely under-utilizing the SLM space–bandwidth product.

In this work, we propose random-phase wave splatting (RPWS), a unified wave optics rendering framework that converts \textit{any 3D representation based on 2D translucent primitives} to random-phase holograms. Our algorithm is fully compatible with recent advances in novel 3D representations based on translucent 2D primitives, such as Gaussians and triangles, improves bandwidth utilization which effectively increases eyebox size, reconstructs accurate defocus blur and parallax, and leverages time-multiplexed rendering not as a mere heuristic for speckle suppression, but as a mathematically exact alpha-blending mechanism derived from first principles in statistics. At the core of RPWS are (1) a fundamentally new wavefront compositing procedure and (2) an alpha-blending scheme specifically designed for arbitrary random-phase geometric primitives, ensuring physically correct color reconstruction and robust occlusion handling when compositing millions of primitives.

RPWS also departs substantially from the recent state-of-the-art primitive-based CGH algorithm, Gaussian Wave Splatting (GWS). Because GWS operates on smooth-phase primitives, it struggles to capture view-dependent effects and realistic defocus blur and severely under-utilizes the SLM space– bandwidth product; moreover, na\"ively extending GWS to random-phase primitives fails to reconstruct accurate colors. In contrast, RPWS is designed from the ground up for arbitrary random-phase translucent primitives, and through extensive simulations and experimental validations we demonstrate that it yields state-of-the-art image quality and perceptually faithful 3D holograms for next-generation near-eye displays.

\end{abstract}




\received{23 May 2025}
\received[revised]{10 August 2025}
\received[accepted]{10 August 2025}

\begin{teaserfigure}
  \centering
	\includegraphics[width=\columnwidth]{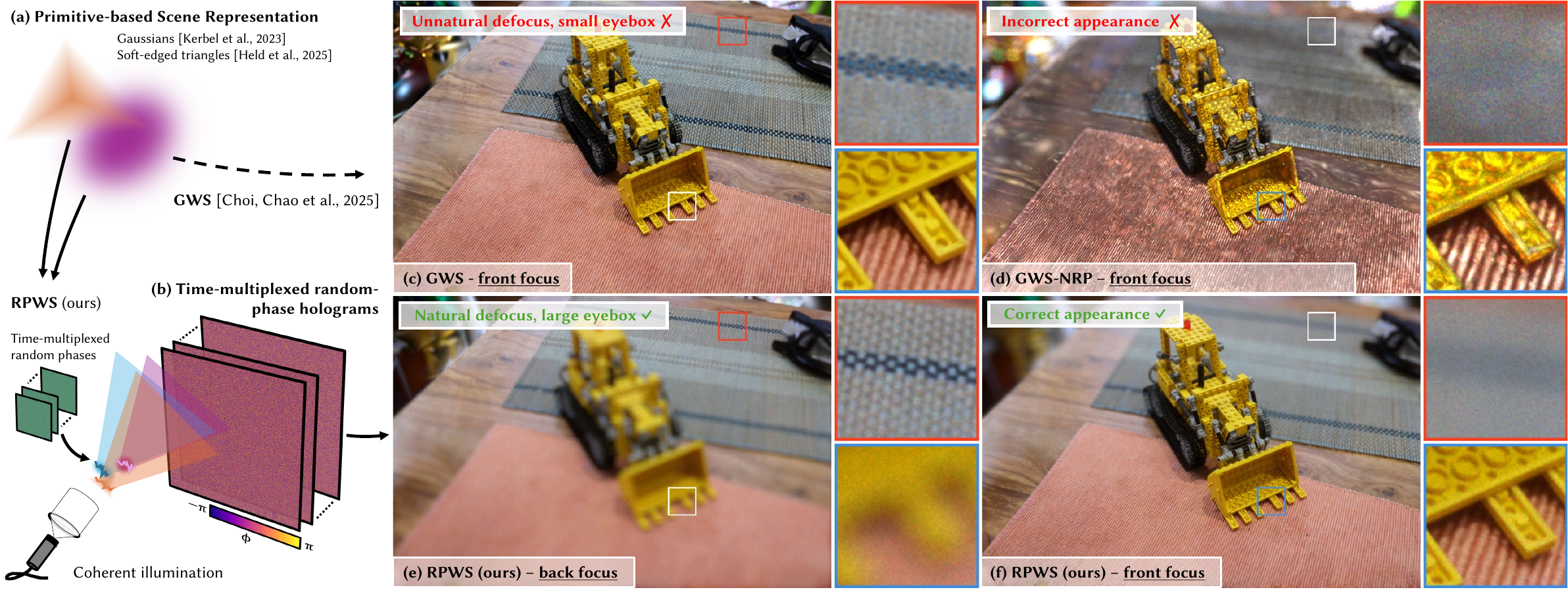}
   \caption{Random-phase Wave Splatting (RPWS) takes in \textit{any 2D primitive-based 3D scene representation}, including Gaussians and triangles, (a) as input and converts it into random-phase holograms (b). The prior state-of-the-art primitive-based CGH method, Gaussian Wave Splatting (GWS) \cite{choi2025gaussian}, is unable to accurately reconstruct natural defocus blur due to the smooth-phase nature of synthesized holograms, resulting in limited blur variation across depths and severe ringing artifacts (c). \textit{N}a\"ively applying \textit{r}andom \textit{p}hase to primitives and running GWS (GWS-NRP), completely fails to reconstruct accurate appearance (d). Our RPWS algorithm, on the other hand, maximally utilizes the bandwidth of the spatial light modulator and reconstructs accurate parallax and defocus blur across the eyebox (e, f). While RPWS works for \textit{arbitrary translucent 2D primitives}, here we show Gaussian-based wave splatting holograms for fair comparison with GWS.}
  \label{fig:teaser}
\end{teaserfigure} 

\maketitle

\section{Introduction}

Holographic near-eye displays offer eyeglasses-like device form factors with unprecedented compactness for virtual and augmented reality display systems~\cite{maimone2017holographic,kim:HolographicGlasses,jang2024waveguide,gopakumar2024full, Choi:2025:HolographicMR}, yet require advanced computer-generated holography (CGH) algorithms to convert a target image or 3D scene into the interference pattern that can be displayed on a spatial light modulator (SLM) \cite{pi2022review,matsushima2020introduction}. Gaussian Wave Splatting (GWS) \cite{choi2025gaussian} recently introduced the idea of direct hologram synthesis from Gaussian primitives \cite{kerbl3Dgaussians, huang20242d}, showing for the first time that neural scene representations can be rendered using wave optics.

However, GWS, as the name suggests, only works with Gaussian primitives, and assumes the phase distribution over the surfaces of the primitives to be near constant or \emph{smooth}, an assumption common in the CGH literature as it typically leads to high in-focus image quality (e.g.,~\cite{maimone2017holographic,shi2021towards}). This fundamentally restricts its ability to reproduce natural defocus, parallax, and view-dependent effects. These simplifications leave a crucial gap: the holographic rendering of random-phase distributions on arbitrary primitive surfaces remains unsolved, even though it is well known that natural reflectance properties of real objects in holography can only be achieved with random phase distributions~\cite{goodman_statistical_optics_2nd,matsushima2020introduction,st1995phase}. Neglecting these phase distributions results in an under-utilization of the inherent space--bandwidth product, or \'etendue, of an SLM, resulting in a small eyebox that leads to poor user experiences~\cite{kim2022accommodative,kim2024holographic}. 


GWS briefly mentioned a high-level sketch in its supplementary materials on how to apply random phase to a \textbf{single Gaussian primitive} to mimic non-diffuse reflectance. However, no strategy was presented for how to \textbf{composite thousands or millions of such random-phase primitives} into a coherent hologram from standard primitive-based scene representations. A na\"ive extension---simply modulating each primitive wavefront with random phase and then running GWS as is (we refer to this \textit{N}a\"ive \textit{R}andom-\textit{P}hase extension as GWS-NRP onwards), which is suggested in the GWS supplementary materials---provably fails, as we demonstrate in Table~\ref{tab:gws_limitations}, Fig.~\ref{fig:teaser}, Section ~\ref{sec:awb_random}, and throughout the paper, resulting in incorrect color reproduction as well as broken parallax and defocus. Moreover, as its name implies, GWS is restricted to Gaussian primitives. How to extend this paradigm to work with other emerging translucent primitives used for novel-view synthesis, such as soft-edged triangles~\cite{sheng20252dtrianglesplattingdirect,burgdorfer2025radianttrianglesoupsoft,Held2025Triangle}, remains unclear. These rapid advances in neural rendering call for a generalizable CGH framework that can convert such primitives into holograms viewable on a 3D display, rather than only producing rendered 2D images.

In this work, we propose Random-phase Wave Splatting (RPWS), a unified computer-generated holography (CGH) framework that converts \textit{3D representations based on translucent primitives}, such as Gaussians~\cite{kerbl3Dgaussians,huang20242d} and soft-edged triangles~\cite{Held2025Triangle,sheng20252dtrianglesplattingdirect,burgdorfer2025radianttrianglesoupsoft}, into random-phase holograms that accurately reconstruct parallax and natural defocus blur.  At the core of RPWS are two key innovations: (1) a novel wave compositing procedure and (2) an intensity-domain alpha blending rule specifically designed to composite millions of random-phase primitives with continuous alpha values across their surfaces. This sharply diverges from prior silhouette-based polygon CGH methods that assume binary alpha values, and it generalizes Gaussian Wave Splatting (GWS) to support arbitrary translucent primitives. Furthermore, we prove that time-multiplexing in RPWS is not a mere heuristic for speckle reduction but the \emph{mathematically exact mechanism} that enables physically correct alpha blending of random-phase wavefronts. These advances collectively lead to high-quality hologram synthesis from state-of-the-art primitive-based 3D scene representations, as we validate through extensive simulation and experimental results of 3D focal stacks and 4D light fields.

\newcolumntype{Y}{>{\centering\arraybackslash}X}

\setlength{\tabcolsep}{4pt} 
\setlength{\textfloatsep}{10pt}
\begin{table}[t!]
    \centering
    \captionsetup{aboveskip=0pt, belowskip=3pt}
    {\footnotesize 
    \caption{\textbf{Limitations of Gaussian Wave Splatting (GWS) \cite{choi2025gaussian}.} GWS ($1^{\text{st}}$ row) generates smooth-phase holograms, resulting in unnatural defocus blur, little-to-no parallax, and small eyebox (1, 3). Na\"ively applying GWS on random-phase primitives ($2^{\text{nd}}$ row, GWS-NRP) results in incorrect color reproduction and erroneous defocus and parallax reconstruction (2, 3). Furthermore, GWS only works with Gaussians, and it is unclear if recent advances in primitive-based representations can also benefit from this wave-splatting paradigm (4). Our random-phase wave splatting (RPWS) framework tackles all these issues. }
    \label{tab:gws_limitations}
    \begin{tabularx}{\linewidth}{m{2.4cm}|YYYY}
        \textbf{Algorithm} & \textbf{1. Large eyebox} & \textbf{2. Correct color} & \textbf{3. Defocus \& parallax} & \textbf{4. Other primitives} \\
        \hline
        GWS \cite{choi2025gaussian} & \redxcell & \greenvcell & \yellowtrcell & \redxcell \\
        \hline
        GWS-NRP & \greenvcell & \redxcell & \yellowtrcell & \redxcell \\
        \hlineB{3}
        RPWS (ours) & \greenvcell & \greenvcell & \greenvcell & \greenvcell \\
        \hline
    \end{tabularx}
    }
\end{table}

Specifically, our contributions include:
\begin{itemize}
    \item Unified \textbf{wave-splatting} (Sec. \ref{subsec:gws_rp}) and \textbf{alpha-blending} algorithms (Sec. \ref{sec:awb_random}) that composite thousands to millions of translucent \textit{random-phase} primitives (including Gaussians and soft-edged triangles), enabling hologram synthesis from a broad class of emerging scene representations.
    \item A rigorous formulation of \textbf{time multiplexing} as the exact alpha-blending mechanism for translucent wavefronts, extending far beyond prior uses as an ad-hoc speckle-reduction heuristic, and statistical optics analysis on \textbf{optimal phase distributions} for accurate random-phase alpha blending (Sec. \ref{sec:awb_random} and supplementary materials).
    \item Extensive experiments, including \textbf{experimental parallax} demonstrations, showing that RPWS achieves superior image quality, perceptually correct defocus blur, and wide parallax reconstructions in both simulation and hardware (Sec. \ref{sec:results}).
\end{itemize}

Source code and example datasets will be made public.

\section{Related Work}
Our work builds on a large body of research on CGH algorithms, which we review below. For a more comprehensive overview of holographic displays, we refer the reader to~\cite{yaracs2010state,park2017recent,chang2020toward,javidi2021roadmap, pi2022review}.

\paragraph{\bf{CGH Algorithms}}
Holograms create a visible image or 3D scene indirectly by displaying an interference pattern, i.e., the hologram, on a 2D amplitude- or phase-only SLM. Methods that convert a target intensity distribution into a hologram are called CGH algorithms. These algorithms have been developed to accommodate a wide variety of input 3D representations, including point clouds~\cite{h2009computer, lucente1993interactive}, meshes~\cite{matsushima2003fast, ahrenberg2008computer}, wireframes~\cite{blinder2021real}, light fields~\cite{padmanaban2019holographic, zhang2019three, park2019non, kang2008accurate, choi2022time}, image layers~\cite{chen2015improved, shi2022end}, and most recently, Gaussians~\cite{choi2025gaussian}. Classic and deep learning–based \textit{direct} CGH methods follow a common pipeline: the target intensity is first encoded into a 2D complex wavefront, typically assuming a certain phase distribution; this wavefront is then propagated to the SLM plane via models such as the angular spectrum method~\cite{goodman2005introduction, pellat1994fresnel}; finally, the resulting complex field is converted into a phase- or amplitude-only pattern, depending on the SLM type~\cite{tsang2013novel, maimone2017holographic, qi2016speckleless}. In contrast, \emph{iterative} CGH methods make use of iterative optimization to achieve a better image quality, albeit at the cost of increased runtime~\cite{gerchberg1972practical, fienup1980iterative, zhang20173d, peng2020neural, chakravarthula2019wirtinger}.

\paragraph{\bf{Phase Distributions of Holograms}}

Although phase is not directly observable, the phase profile of a wavefront plays a crucial role in determining the spatio-angular behavior of the observable light field~\cite{st1995phase, schiffers2023stochastic, kim2022accommodative, Chakravarthula2022pupil}. For this reason, two popular heuristics have been developed that are widely used in  CGH literature: \textit{smooth-phase} and \textit{random-phase} holograms~\cite{maimone2017holographic, yoo2021optimization}. Smooth-phase, sometimes called random-phase-free, holograms~\cite{shimobaba2015random} achieve high image quality that can be demonstrated with relatively simple experimental setups~\cite{maimone2017holographic, shi2021towards, peng2020neural, choi2025gaussian}. The main drawback of smooth-phase distributions, however, is that they concentrate energy in the low frequencies of the angular spectrum, resulting in a severely restricted eye box size, limited defocus effects, and increased sensitivity to pupil position. These effects limit the perceptual realism and overall user experience of the produced holograms~\cite{kim2024holographic} as well as the support for perceptually important focus cues~\cite{kim2022accommodative}.

On the other hand, random-phase holograms are capable of reconstructing larger parallax and natural defocus blur, which is necessary for a perceptually realistic and natural viewing experience \cite{kim2024holographic}. However, rapid phase variations between adjacent pixels introduce unwanted speckle noise created by constructive and destructive interference~\cite{goodman2007speckle}. Speckle reduction techniques often utilize some form of partial coherence, introduced by partially coherent or multiple coherent light sources, or more commonly through time multiplexing~\cite{lee2020light, chao2024large, kuo2023multisource, choi2022time, curtis2021dcgh, peng2021speckle}. 

In this paper, we prove that only a specific family of random-phase distributions yields correct alpha blending of wavefronts (Sec. \ref{sec:awb_random}). We then show that the conventional uniform $2\pi$ distribution used in prior random-phase literature conveniently happens to be one of them, allowing our RPWS holograms to maximally utilize the SLM bandwidth for large parallax and natural defocus reconstruction. Finally, we show that time multiplexing is not only a speckle reduction heuristic for random-phase holography, but the statistically exact mechanism that enables accurate alpha blending.

\paragraph{\bf{Primitive-based 3D Representations and CGH Algorithms.}} Among the various CGH algorithms described above, polygon-based CGH algorithms that use meshes as the input 3D format are closely related to our approach \cite{matsushima2005computer, matsushima2005exact, matsushima2009extremely, askari2017occlusion, chen2009computer, matsushima2018full, matsushima2014silhouette}. However, the constraint that each triangle has to be fully opaque limits the expressivity of mesh models by preventing soft blending of primitives, a paradigm central to recent 3D reconstruction frameworks such as 3DGS \cite{kerbl3Dgaussians}. Coupled with per-primitive texturing, these restrictions make polygon-based CGH difficult to capture fine details without using an excessive number of tiny triangles. Gaussian Wave Splatting \cite{choi2025gaussian}, on the other hand, has emerged as a new primitive-based CGH algorithm that converts Gaussian-based scene representations \cite{kerbl3Dgaussians, huang20242d} to complex holograms, greatly outperforming traditional polygon-based CGH in terms of image quality and rendering time. However, GWS does not support view-dependent effects and natural defocus due to its smooth-phase nature. Additionally, the alpha-blending and wave-splatting procedure described in GWS does not naturally extend to random-phase Gaussians. Na\"ively applying GWS to random-phase wavefronts, as suggested in the GWS supplementary materials (which we refer to as GWS-NRP), provably fails as we demonstrate throughout the paper. Finally, GWS solely works with Gaussians, thus it remains unknown if such a wave compositing scheme can be extended to support a larger variety of translucent geometric primitives widely used in recent 3D reconstruction literature, such as soft-edged triangles \cite{Held2025Triangle, sheng20252dtrianglesplattingdirect, burgdorfer2025radianttrianglesoupsoft}.

In this work, we show that for random-phase primitive wavefronts, alpha blending is exact \textit{in expectation} in the \textit{intensity domain}, in contrast to the amplitude domain for smooth-phase primitives as described in GWS (Sec. \ref{sec:awb_random}). We then devise a novel wavefront composition technique that specifically works with arbitrary random-phase geometric primitives (Sec. \ref{subsec:gws_rp}) for accurate color reconstruction and occlusion handling. As such, our algorithm generates holograms with high image quality by fully leveraging the translucent primitive representations while synthesizing accurate defocus blur and wide parallax via random phase and time-multiplexing.

\section{Wave-optics Rendering of Translucent Primitives for Computer-generated Holography}
Our approach takes as input a set of multi-view images that are turned into \textit{translucent} 2D primitives (primitives with continuous alpha values ranging from 0 to 1) representing a 3D scene. These primitives are then converted to time-multiplexed holograms through random-phase sampling. We briefly review the relevant background on emerging primitive-based scene representations (e.g. Gaussian splats \cite{kerbl3Dgaussians, huang20242d}) and existing work on \emph{smooth-phase} Gaussian wave splatting, before introducing our \emph{random-phase} wave splatting algorithm, which uniquely leverages a time-multiplexed image formation for accurate alpha blending of arbitrary geometric primitives. 


\subsection{Background}
\label{sec:background}

\subsubsection{Primitive-based Scene Representations}

Since the advent of 3DGS \cite{kerbl3Dgaussians}, significant research has focused on developing alternative geometric primitives for 3D scene reconstruction.
Among these emerging primitive-based representations, 2D translucent primitives are notable for their local manifold structures, which enable explicit normal computation and surface extraction. Their flat-surface formulation integrates naturally with rasterization pipelines and aligns well with silhouette-based methods for polygonal CGH, allowing us to draw inspiration from prior work \cite{matsushima2003fast, matsushima2005computer, matsushima2014silhouette, matsushima2020introduction}. 

Examples of such 2D translucent primitives include 2D Gaussians \cite{huang20242d} and soft-edged triangles \cite{Held2025Triangle, sheng20252dtrianglesplattingdirect, burgdorfer2025radianttrianglesoupsoft}. For 2D Gaussians, the geometry of each of these $i=1 \ldots N$ Gaussians is described by its mean $\mean_i \in \mathbb{R}^3$, 3D covariance $\cov_i=\rot_i \sca_i \sca_i^\mathsf{T} \rot_i^\mathsf{T} \in \mat{3}$ that can be factorized into a rotation matrix $R_i \in \mat{3}$ and a scaling matrix $S_i \in \mat{3}$. For triangles \cite{Held20243DConvex}, the geometry of the $i^{\text{th}}$ triangle is defined by its three vertices $\mathbf{V}_i \in \mat{3}$ and a smoothness factor $\sigma_i\in \mathbb{R}$ that describes the transparency falloff from the edges of the triangle. All of these primitives typically also include two additional parameters: opacity $o_i \in \mathbb{R}$ that determines its global transparency and color $c_i \in \mathbb{R}$ (for a single color channel). Please refer to these papers for more details on the definitions of the primitives.

Any primitive-splatting approach requires the $N$ primitives representing a scene to be depth sorted from \textit{front to back} based on the $z$ value with respect to the camera position, or in a holographic display setup, the SLM plane. We closely follow the \textit{holographics} pipeline described in GWS \cite{choi2025gaussian} that transforms these primitives into an adequate hologram space for CGH calculation.

\subsubsection{Gaussian Wave Splatting}
Gaussian Wave Splatting (GWS) \cite{choi2025gaussian} is a CGH method capable of computing holograms that accurately represent 3D scenes from collections of 2D Gaussians \cite{huang20242d} extracted from any off-the-shelf optimized 2DGS models, e.g., models optimized using the \texttt{gsplat} library~\cite{ye2024gsplat}. GWS first analytically determines the spectrum of each Gaussian described by its mean $\mean_i$ and 3D covariance $\cov_i$ and computes the wavefront  $u_i(\coordinate)=a_i(\coordinate)e^{ikz_i}$, where $z_i = (\mean_i)_z$ is the Gaussian object depth and $k = \frac{2\pi}{\lambda}$. Then, each wavefront $u_i(\coordinate)$ is propagated using the angular spectrum propagation operator $\prop(\cdot; z)$~\cite{goodman2005introduction, matsushima2009band} and alpha blended from \textit{front to back} using the opacity $o_i$ and color $c_i$ associated with each Gaussian to get the final composited wavefront profile at the SLM, given by Eqs.~\ref{eq:gwsplat} and \ref{eq:gwsplat_T}. We refer to this process as \textit{alpha wave blending}:

\begin{align}
    u_{\textrm{SLM}} (\coordinate) = \sum_{\idx}^\maxidx \mathcal{P}\Big(c_\idx o_\idx |u_\idx(\coordinate)|\transmittance_\idx\left(\coordinate\right) e^{ikz_i}; -z_\idx\Big), \label{eq:gwsplat}\\
    \transmittance_\idx\left(\coordinate\right) = \prod_{j=1}^{\idx-1} (1 - o_{j} |u_j(\coordinate)|).
    \label{eq:gwsplat_T}
\end{align}

GWS inherits the ability of Gaussian splatting to seamlessly merge large numbers of Gaussians for high-quality reconstruction. GWS further collapses classic alpha blending and volume rendering \cite{kerbl3Dgaussians, kajiya1984ray} if we ignore the wave propagation operator and match the phase of all wavefronts at all depths (i.e., $\angle u_\idx=kz_i$) such that the composited wavefront at the SLM plane achieves a smooth or near-constant phase profile, which we will formally derive in Section \ref{sec:awb_random}. Therefore, this formulation of GWS inherently generates ``smooth-phase'' holograms. Although GWS has demonstrated the potential to recreate sharp details with photorealistic image quality ~\cite{choi2025gaussian}, prior works in holography \cite{schiffers2023stochastic, kim2022accommodative, shi2024ergonomic, lee2022binary} have pointed out that smooth-phase holograms are undesirable due to their poor SLM bandwidth utilization, unnatural defocus blur, large depth of field (i.e., small blur variation across different depths), and floater artifacts. GWS also requires $|u_i(\coordinate)|$ to be a 2D Gaussian distribution, thus its generalizability to other geometric primitives remains unknown.

Most importantly, na\"ively extending GWS to random-phase primitives (GWS-NRP) by additionally modulating each Gaussian wavefront with a random phase map $\phi_i \overset{\text{iid}}\sim \mathcal{U}(-\pi, \pi)$ as shown below:
\begin{align}
    u_{\textrm{SLM}} (\coordinate) = \sum_{\idx}^\maxidx \mathcal{P}\Big(c_\idx o_\idx |u_\idx(\coordinate)|\transmittance_\idx\left(\coordinate\right) e^{ikz_i + \phi_i}; -z_\idx\Big)
    \label{eq:gwsplat_naive_random}
\end{align}
provably fails as we demonstrate in Section ~\ref{sec:awb_random}.

\subsection{Random-phase Wave Splatting}
\label{subsec:gws_rp}

 To formally close this gap, we propose \textit{Random-phase Wave Splatting} (RPWS), a robust CGH framework that generates random-phase holograms from translucent primitives, tackling all of the drawbacks of GWS. In RPWS, each primitive wavefront is effectively modulated by a random phase map $\phi_i(\coordinate)$ to scatter the light passing through each primitive in the scene away from the optical axis. The primitives are then alpha blended and composited from \textit{back to front} with respect to the SLM, and multiple such wavefronts with individually sampled random phases are time multiplexed for accurate alpha blending to achieve the desired intensity distribution. The novel alpha blending and wave compositing procedure in RPWS is specifically designed to work with random-phase \textit{translucent} primitives with continuous alpha values, and greatly outperforms GWS in blending complex wavefronts \cite{choi2025gaussian}, which we demonstrate in Sec. \ref{sec:results}.


In Eq. ~\ref{eq:gwsplat_random_tm}, let $g_i(\coordinate)$ denote the \textit{back-to-front} composited wavefront starting from the $N^{th}$ primitive up to the $i^{th}$ primitive and $M_i(\coordinate)$ be the transmittance mask of the $i^{th}$ primitive. The primitive wavefront \( u_i(\coordinate) \) denotes the wavefront generated from primitives such as Gaussians or translucent triangles (Section~\ref{sec:results}), making RPWS a plug-and-play framework for arbitrary translucent wavefronts. The wavefront at the ${i - 1}^{\text{th}}$ parallel plane where the next primitive is located, which is a propagation distance $\Delta z = z_{i - 1} -  z_{i}$ away, is given by:

\begin{align}
    \hspace*{-0.5em}
    g_{i - 1}^{(t)} (\coordinate) = \mathcal{P} \Big( M_i(\coordinate) \; g_i(\coordinate) +  \sqrt{c_i} \sqrt{ o_i |u_i(\coordinate)|} \; e^{i \angle u_i(\coordinate) + \phi_i^{(t)}(\coordinate)} ; \Delta z\Big)
    \label{eq:gwsplat_random_tm}, \\
    M_i(\coordinate) = \sqrt{(1 - o_{i} |u_i(\coordinate)|)},
    \label{eq:gsplat_mask}
\end{align}
where $t, 1\leq t\leq T$ is the index of the time-multiplexed frame, $T$ is the total number of multiplexed frames, and $\phi_i^{(t)}$ is the sampled random phase for the $i^\text{th}$ primitive at the $t^\text{th}$ frame. The final composited wavefront at the SLM plane that is located at $z_0 = 0$ is simply defined by $u_\text{SLM}^{(t)}(\coordinate) = g_0^{(t)}(\coordinate)$. We highlight that the above procedure is highly similar to the \textit{over} operation \cite{porter1984over}, or Painter's algorithm, in traditional graphics, with additional wave propagation and square root operations. This is because in RPWS, we use primitives optimized to reconstruct the \textit{square} (i.e., intensity) of the target scene, instead of the amplitude like in GWS, which we explain in detail in the following Section~\ref{sec:awb_random}.

The intensity of reconstructed images of the time-multiplexed hologram is flexibly described by the operator $\mathcal{O(\cdot; \cdot)}$ as
\begin{align}
    I(\mathbf{x})  = \frac{1}{T} \sum_{t=1}^{T} \left| \mathcal{O} \left( u_\text{SLM}^{(t)}(\mathbf{x}); \cdot \right) \right|^2 ,
\end{align}
where the operator $\mathcal{O(\cdot; \cdot)}$ could describe a single propagation that reconstructs a 2D image, multiple propagations that reconstruct a 3D focal stack, or the Short-time Fourier Transform (STFT) that reconstructs a 4D light field \cite{choi2022time}.


With random-phase modulation, combined with the novel wave compositing and alpha blending schemes, the bandwidth of the SLM can be maximally utilized while enabling accurate reconstruction of occlusion and color. This leads to a large eyebox and parallax, shallow depth of field (i.e., significant blur variation across depths), and natural defocus blur, as illustrated in Section~\ref{sec:results}. Please refer to the supplemental materials for a detailed statistical-optics analysis quantifying bandwidth and defocus.

\subsection{Random-phase Alpha Wave Blending}
\label{sec:awb_random}

RPWS departs fundamentally from GWS~\cite{choi2025gaussian} and prior silhouette-based methods~\cite{matsushima2014silhouette} in its treatment of alpha blending. While earlier smooth-phase CGH approaches assumed \textbf{amplitude}-domain alpha compositing, we show that for random-phase wavefronts alpha blending is linear in the \textbf{intensity} domain \textit{in expectation}. This crucial distinction, which is largely overlooked in prior work and GWS, directly leads to our wave compositing equation (Eq.~\ref{eq:gsplat_mask}), where square-root terms and time-multiplexing naturally emerge, in contrast to GWS. We analyze different alpha blending formulations in detail in Fig. \ref{fig:alpha_blending_comparison} and in Section \ref{subsubsec:alpha_blending}. 

For simplicity, consider two primitive wavefronts at the same depth,
$u_1 = \alpha_1 e^{i\phi_1},\;u_2 = \alpha_2 e^{i\phi_2} \in \mathbb{C}^2$,
with $u_1$ in front of $u_2$. Here $\alpha_1, \alpha_2 \in \mathbb{R}^2$ are alpha maps defined as $\alpha = o \cdot \mathcal{M}$, the product of the primitive opacity $o \in \mathbb{R}$ and its transparency falloff $\mathcal{M} \in \mathbb{R}^2$ (from the Gaussian covariance/mean or the $\sigma$ parameter of a translucent triangle~\cite{Held2025Triangle}). Let $c_1, c_2 \in \mathbb{R}$ denote the scalar colors. In standard ray-based splatting~\cite{kerbl3Dgaussians, huang20242d, Held2025Triangle}, the alpha-blended color is
\begin{align}
c = \alpha_1 c_1 + (1 - \alpha_1)\alpha_2 c_2,
\label{eq:gs_ab}
\end{align}
where $c$ is the resulting amplitude of the blended wavefront.

In wave splatting, the key question is how to alpha-blend primitive \textit{wavefronts} in the presence of phase. Formally, this requires finding blending weights $w_1(\alpha_1, \alpha_2, c_1, c_2), w_2(\alpha_1, \alpha_2, c_1, c_2) \in \mathbb{R}$ such that the amplitude of the blended wavefront  
\begin{align}
    u = w_1(\alpha_1, \alpha_2, c_1, c_2) e^{i\phi_1} + w_2(\alpha_1, \alpha_2, c_1, c_2) e^{i\phi_2}
    \label{eq:blended_wavefront}
\end{align}
matches the target color $c$ reconstructed from the input primitives.

Squaring the amplitude of both sides of Eq.~\ref{eq:blended_wavefront} yields
\begin{align}
    c^2 = |u|^2 &= \big| w_1 e^{i\phi_1} + w_2 e^{i\phi_2} \big|^2 \\
          &= w_1^2 + w_2^2 + w_1 w_2 \left( e^{i(\phi_1 - \phi_2)} + e^{i(\phi_2 - \phi_1)} \right) \\
          &= w_1^2 + w_2^2 + 2w_1 w_2 \cos(\phi_1 - \phi_2),
    \label{eq:awb_random}
\end{align}
which shows explicit dependence on the phase distribution of $\phi_1 - \phi_2$, as analyzed below.

\begin{figure}
    \centering
    \includegraphics[width=\linewidth]{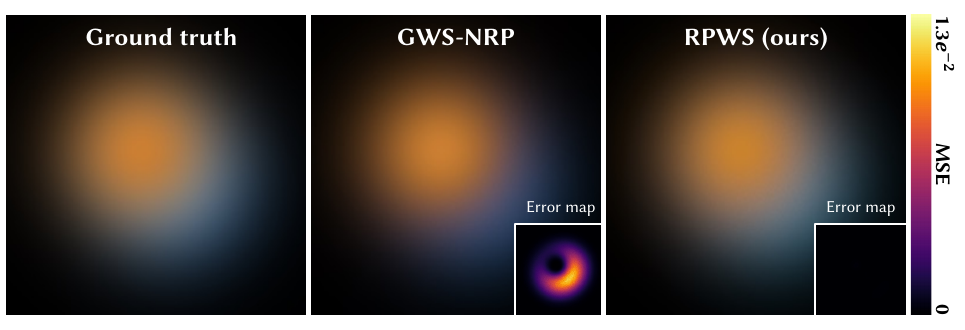}
    \caption{\textbf{Effectiveness of RPWS sqrt-blending of random-phase wavefronts.} We show the effectiveness of our sqrt-blending procedure using the toy example described in Section \ref{sec:awb_random} with two random-phase Gaussians, where the orange Gaussian is in front of the blue Gaussian and alpha blended. The straightforward extension of GWS amplitude-based blending with random phase (GWS-NRP) fails to reconstruct the correct appearance with significant errors near occlusion borders --- which is exactly the cross term in Eq. \ref{eq:awb_random}. RPWS sqrt-blending achieves perfect alpha blending results.}
    \label{fig:placeholder}
\end{figure}

\subsubsection{Smooth-phase (GWS): $\phi_1 - \phi_2 = 2k\pi$}
When two primitive wavefronts are in phase at the same depth, $\cos(\phi_1-\phi_2)=1$, and Eq.~\ref{eq:awb_random} reduces to $(w_1+w_2)^2$, i.e., $|u|=c=w_1+w_2$. Comparing with Eq.~\ref{eq:gs_ab}, we set $w_1=\alpha_1 c_1$ and $w_2=(1-\alpha_1)\alpha_2 c_2$, yielding
\begin{align}
    u = \alpha_1 c_1 e^{i\phi_1} + (1-\alpha_1)\alpha_2 c_2 e^{i\phi_2}.
    \label{eq:awb_smooth}
\end{align}
which is linear in the amplitude domain, exactly the \textit{alpha wave blending} formulation described in GWS.

\subsubsection{Random-phase (RPWS): $\phi_1, \phi_2 \overset{\text{iid}}\sim \mathcal{U}(-\pi, \pi)$}
When primitives are modulated by i.i.d. random phases drawn from $\mathcal{U}(-\pi,\pi)$, the cross term $\cos(\phi_1-\phi_2)$ in Eq.~\ref{eq:awb_random} is nonzero for a single realization, but only vanishes \textit{in expectation}, which is essentially time multiplexing. Specifically, the expected intensity is
\begin{align}
    \mathbb{E}[|u|^2] &= w_1^2 + w_2^2 + 2w_1w_2\mathbb{E}[\cos(\phi_1-\phi_2)] \label{eq:cross_term} \\ 
    &= w_1^2 + w_2^2 = c^2,
    \label{eq:awb_random_2}
\end{align}
showing that cross terms are completely eliminated in the intensity domain.

To satisfy Eq.~\ref{eq:awb_random_2}, we optimize primitives with the standard ray-based alpha blending rule on intensity $c^2$,
\begin{align}
    c^2 = \alpha_1'c_1' + (1-\alpha_1')\alpha_2'c_2',
\end{align}
and set
\begin{align}
    w_1 = \sqrt{\alpha_1'c_1'}, \quad 
    w_2 = \sqrt{(1-\alpha_1')\alpha_2'c_2'}.
\end{align}
The resulting random-phase compositing rule is
\begin{align}
    u = \sqrt{\alpha_1'c_1'} \, e^{i\phi_1} + 
        \sqrt{(1-\alpha_1')\alpha_2'c_2'} \, e^{i\phi_2},
    \label{eq:awb_random_final}
\end{align}
where each primitive wavefront is assigned color $\sqrt{c_i'}$ from a model optimized to reconstruct $c^2$ (intensity) rather than $c$ (amplitude). This formulation generalizes directly to thousands or millions of random-phase primitives (Section~\ref{subsec:gws_rp}), forming the basis of our RPWS wave-compositing algorithm in Eq. \ref{eq:gwsplat_random_tm} and \ref{eq:gsplat_mask}. 

Evident from Eq. \ref{eq:awb_random_2}, we emphasize that time multiplexing is not merely an optional add-on for speckle reduction, but a fundamental requirement for correct alpha blending in our framework. The alpha blending equation is inherently stochastic: it is only satisfied in expectation over random phases, and any single realization introduces cross-term interference that corrupts the blend. Time multiplexing ensures that these cross terms vanish statistically, thereby realizing the mathematically correct alpha blending behavior in practice. In this sense, speckle reduction is only a by-product; the primary role of time multiplexing is to enforce the validity of the probabilistic alpha blending formulation itself. We further derive exact convergence bounds in the supplemental materials, showing that the uniform $2\pi$ phase distribution belongs to a class of phase distributions that achieve the optimal $O(1/\sqrt{N})$ decay of cross terms, thereby providing formal justification for our probabilistic alpha blending framework.

\subsection{Applying Random Phase to Primitives}
To apply random phase to primitive surfaces, we adopt the heuristic briefly mentioned in the supplementary materials of GWS \cite{choi2025gaussian}. While GWS only alluded to this idea qualitatively, we provide, \textbf{for the first time}, a rigorous mathematical formulation grounded in statistical optics that proves this heuristic to be exact, and demonstrate its practical applications such as programmatic depth-of-field control. Please refer to the supplementary materials of this paper for more details.

\section{Experiments}
\label{sec:results}
\subsection{Implementation Details}
We generate 3D holograms of scenes from the Blender~\cite{mildenhall2020nerf} and MipNeRF-360~\cite{barron2022mipnerf360} datasets using 2D Gaussians, textured meshes, and translucent triangles. Following GWS~\cite{choi2025gaussian}, we prepare 3D scene representations with the open-source \texttt{gsplat} library~\cite{ye2024gsplat} for 2DGS, NeRF2Mesh models~\cite{tang2022nerf2mesh} for textured meshes, and Triangle Splatting models~\cite{Held2025Triangle} for translucent triangles. Additional data preparation details are provided in the supplemental materials.

We implement our algorithms in PyTorch and build a benchtop holographic display for experimental validation. Please refer to the supplementary materials for detailed pseudocode and hardware specifications.

\subsection{Simulation and Experimental Results}
\label{subsec:results}
\subsubsection{Baseline Comparisons with Simulation Results. }

We compare RPWS against several primitive-based CGH baselines, including random-phase polygon-based CGH (Polygons-RP) via the silhouette method~\cite{matsushima2003fast, matsushima2005exact, matsushima2009extremely} and its time-multiplexed variants (1 and 8 frames), as well as Gaussian Wave Splatting (GWS)~\cite{choi2025gaussian}. For all quantitative and qualitative baseline comparisons, we restrict RPWS to 2DGS representations (RPWS-GS), where the number of Gaussians can be matched directly to the number of polygon primitives (using 3DGS-MCMC \cite{kheradmand20243dgsmcmc}) for fair comparisons, similar to GWS. For RPWS with soft-edged triangle primitives \cite{burgdorfer2025radianttrianglesoupsoft, Held2025Triangle, sheng20252dtrianglesplattingdirect}, which we denote as RPWS-$\Delta$, we pick triangle splatting \cite{Held2025Triangle} as the representative method. Since triangle splatting does not allow for exact primitive count control, we only show qualitative results of RPWS-$\Delta$.

\begin{figure*}[ht!]
    \centering
    \includegraphics[width=0.96\textwidth]{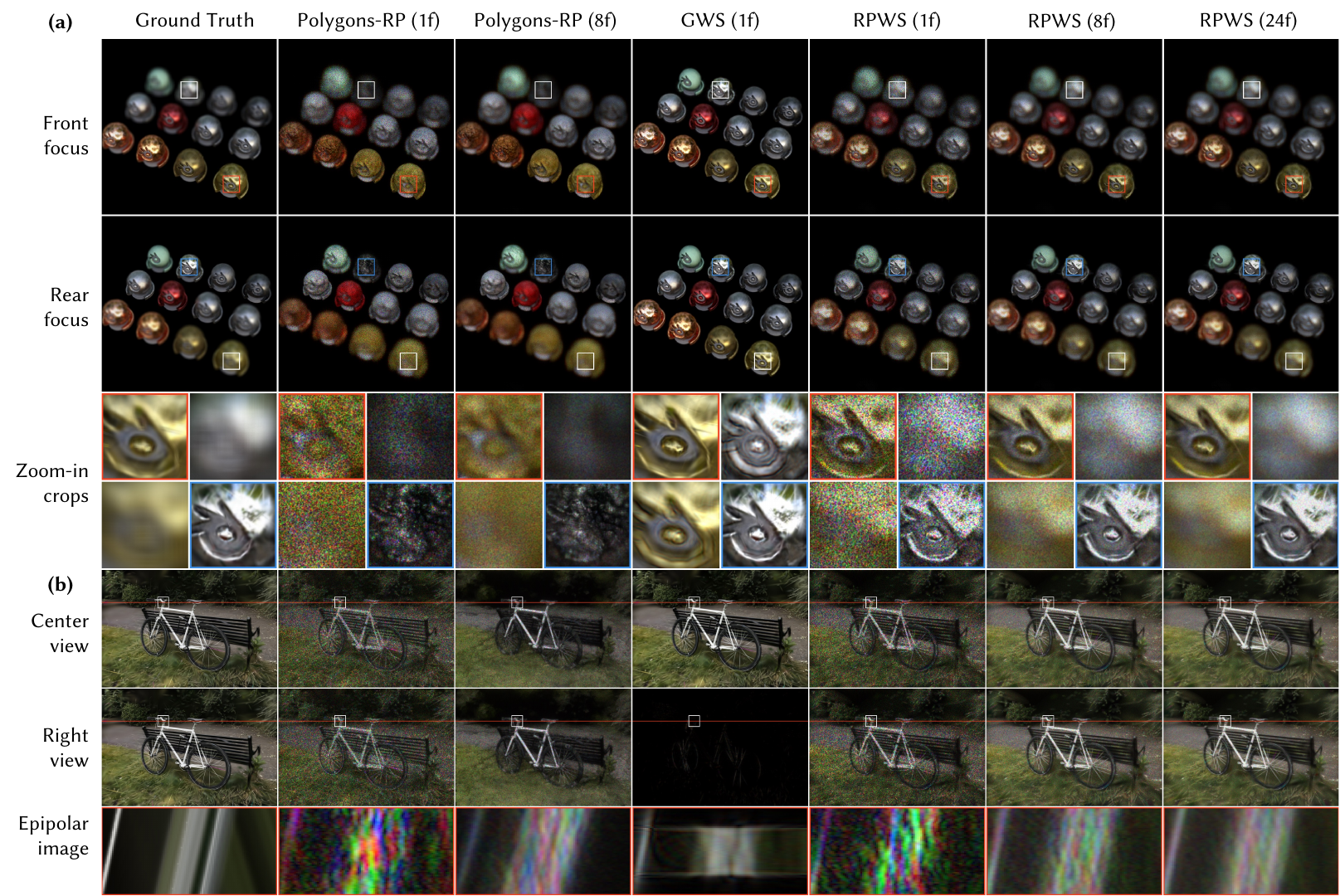}
    \caption{\textbf{Simulated 3D focal stacks and 4D light fields reconstructed from various baseline CGH algorithms.} The image quality of random-phase polygons-based CGH (Polygons-RP) is inherently limited by the coarse per-face textured mesh representation, resulting in poor image quality even in in-focus regions. GWS \cite{choi2025gaussian} reconstructs sharp details at in-focus regions, but suffers from large depth of field and unnatural ringing artifacts. Our method (RPWS) generates sharp content at focused regions and the resulting hologram has shallow depth of field, reconstructing natural defocus blur across different depths. With additional time-multiplexing, the image quality of RPWS significantly improves.}    
    \label{fig:sim_fs}
    
\end{figure*}

\begin{figure*}[ht!]
    \centering
    \includegraphics[width=0.96\textwidth]{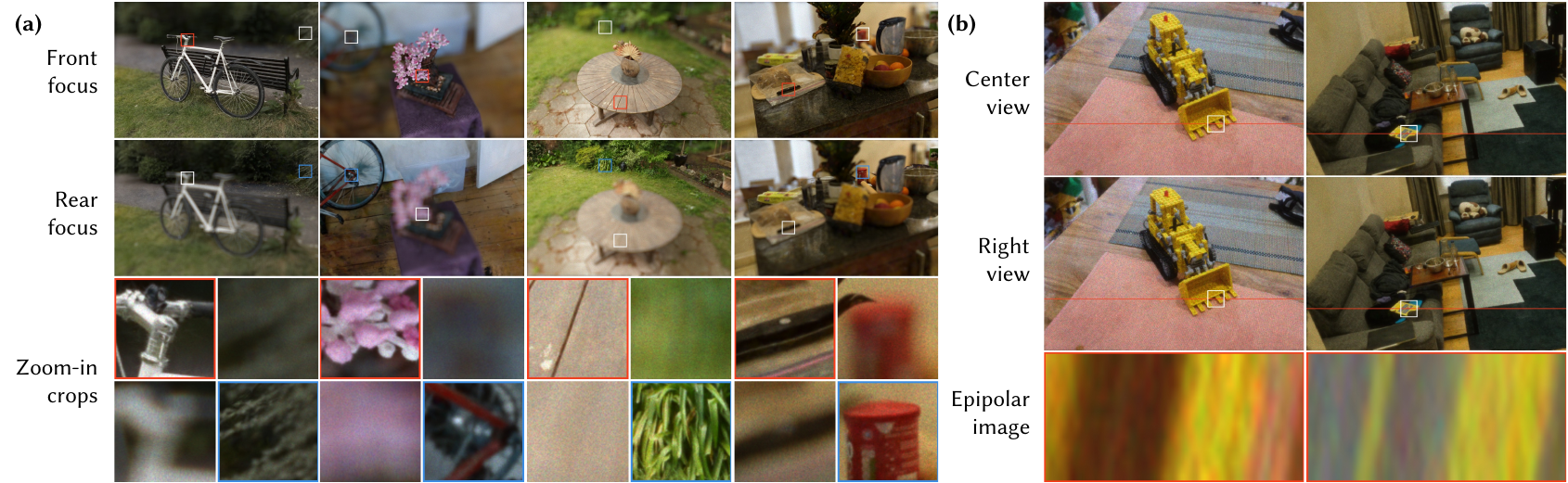}
    \caption{\textbf{Simulated 3D focal stacks and 4D light fields reconstructed from RPWS of triangle splats \cite{Held2025Triangle}. } We run our RPWS algorithm on Triangle splats proposed by \citet{Held2025Triangle}. RPWS with triangle splats accurately reconstructs natural defocus blur and parallax on a wide variety of scenes, validating the robustness of our method to different translucent primitive types. }
    \label{fig:triangle_splats}
\end{figure*}

Fig.~\ref{fig:sim_fs}-(a) and ~\ref{fig:triangle_splats}-(a) compare simulated focal stacks of the synthesized holograms. GWS produces smooth-phase holograms with little depth-dependent variation, yielding unnatural coherent blur with ringing artifacts. Single-frame random-phase holograms (Polygon-RP, RPWS) exhibit strong speckle, but quality improves with time multiplexing (8, 24 frames). Polygon-RP produces more natural defocus blur, but its in-focus fidelity is limited by the low-quality per-face textured mesh representation. Both RPWS-GS and RPWS-$\Delta$ achieves the most natural defocus blur, closely matching incoherent blur, while maintaining in-focus quality comparable to GWS.

\begin{table}[t!]
   \footnotesize
   \centering
   \renewcommand{\arraystretch}{1.2} 
   \begin{tabular}{@{}P{0.3\linewidth}P{0.31 \linewidth}P{0.31 \linewidth}@{}}
   \toprule
   \makecell[l]{CGH Algorithm} &
   Blender &
   Mip-NeRF 360 \\
   \midrule
   \makecell[l]{\scalebox{0.95}{Polygons-RP (1 frame)} \\ \cite{matsushima2005computer}} & 17.02 / 0.14 / 0.75 & 12.77 / 0.14 / 0.76 \\
   \makecell[l]{\scalebox{0.95}{Polygons-RP (8 frames)} \\ \cite{matsushima2005computer}} & 20.27 / 0.30 / 0.65 & 15.63 / 0.27 / 0.69 \\
   \makecell[l]{\scalebox{0.95}{RPWS-GS (1 frame)}} & 18.11 / 0.15 / 0.76 & 14.17 / 0.17 / 0.76 \\
   \makecell[l]{\scalebox{0.95}{RPWS-GS (8 frames)}} & 23.42 / 0.36 / 0.64 & 19.65 / 0.36 / 0.65 \\
   \makecell[l]{\scalebox{0.95}{RPWS-GS (24 frames)}} & \textbf{24.87} / \textbf{0.49} / \textbf{0.56} & \textbf{21.21} / \textbf{0.48} / \textbf{0.58} \\
   \noalign{\vskip 0.5ex}
   \hline
   \noalign{\vskip 0.5ex}
    \makecell[l]{\scalebox{0.95}{GWS \cite{choi2025gaussian}}} & 14.35 / 0.18 / 0.63 & 10.50 / 0.17 / 0.60 \\
   \bottomrule
   \end{tabular}
\caption{\textbf{Quantitative light-field reconstruction performance of different CGH algorithms. } We evaluate the image quality of $10\times10$ dense light fields reconstructed from the simulated holograms generated using different CGH methods in terms of PSNR $(\uparrow)$ / SSIM $(\uparrow)$ / LPIPS $(\downarrow)$. The best performing metrics for all CGH baselines are boldfaced. Our method achieves the best image light-field reconstruction performance and eyebox uniformity among all CGH baselines.}
\label{tab:quan_metrics_lf}
\end{table}

\begin{figure*}[!ht]
    \centering
    \includegraphics[width=\textwidth]{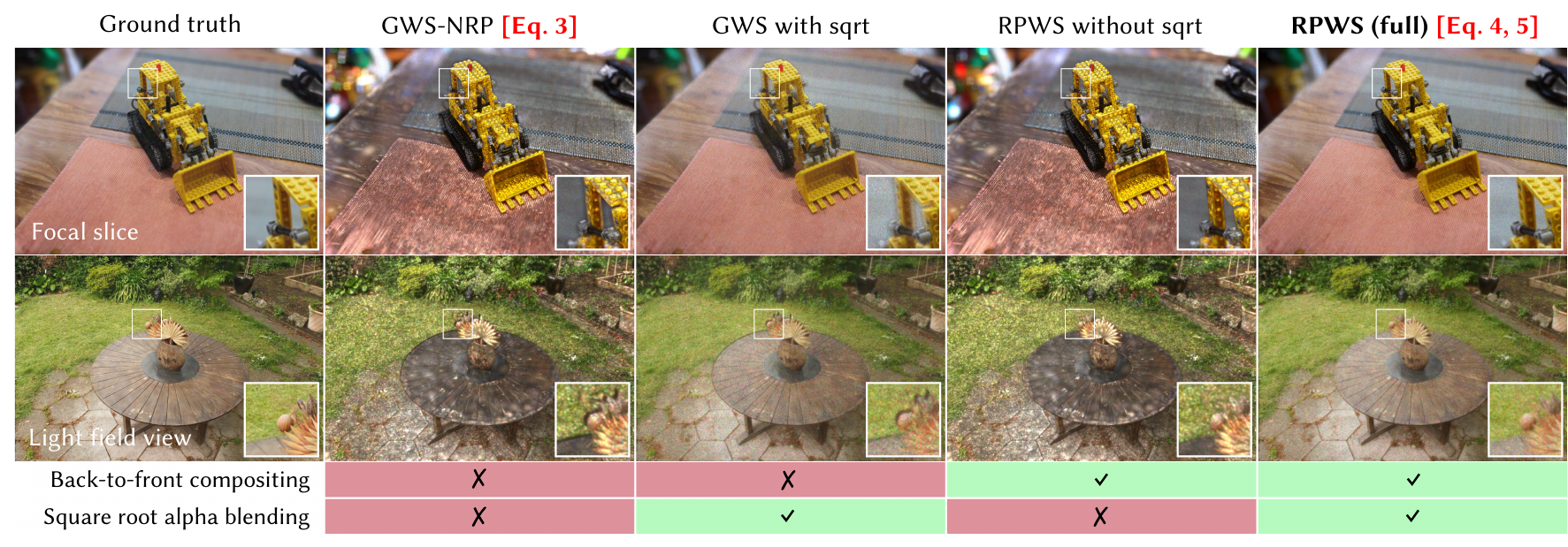}
    \caption{\textbf{Ablation study on random-phase CGH rendering algorithms. } We compare our RPWS algorithm with three other alpha blending methods on random-phase primitives. (a) Na\"ively applying random phase to GWS fails to reconstruct accurate appearance. (b) GWS with our square root alpha blending formulation generates somewhat accurate appearance. However, incorrect occlusion handling at depth discontinuities manifests as light leakage and halo artifacts in focal stacks and black holes near occlusion borders. This is an artifact frequently observed in conventional multifocal displays \cite{narain2015multi, mercier2017fast, chang2020toward}. (c) RPWS without square root-based blending also fails to reconstruct accurate colors. (d) Our full RPWS model reconstructs accurate color, natural defocus blur, and physically-correct parallax. Please refer to the supplementary materials for equations and pseudocodes of all alpha-blending formulations.}
    \label{fig:alpha_blending_comparison}
\end{figure*}

Fig.~\ref{fig:sim_fs}-(b) and ~\ref{fig:triangle_splats}-(b) compare simulated light fields of the synthesized holograms using individual views and epipolar images. For smooth-phase GWS, wavefront energy concentrates near the eyebox center, causing severe degradation at the periphery. In contrast, RPWS-GS and RPWS-$\Delta$ distribute energy evenly across the eyebox, enabling accurate view reconstruction at all pupil positions. Random-phase polygon-based CGH again suffers from the coarse per-face textured mesh representation. Quantitative results for dense $10 \times 10$ light-field reconstructions, measured with standard image quality metrics, are reported in Table~\ref{tab:quan_metrics_lf} where RPWS clearly outperforms all methods.

Despite belonging to fundamentally different algorithmic classes, we delineate the key distinctions between RPWS and STFT-based light field CGH methods \cite{choi2022time, kim2024parallax} and perform additional baseline comparisons in the supplementary materials.

\subsubsection{Experimentally Captured Focal Stack and Parallax Results. } 
\begin{figure*}[ht!]
    \centering
    \includegraphics[width=0.95\textwidth]{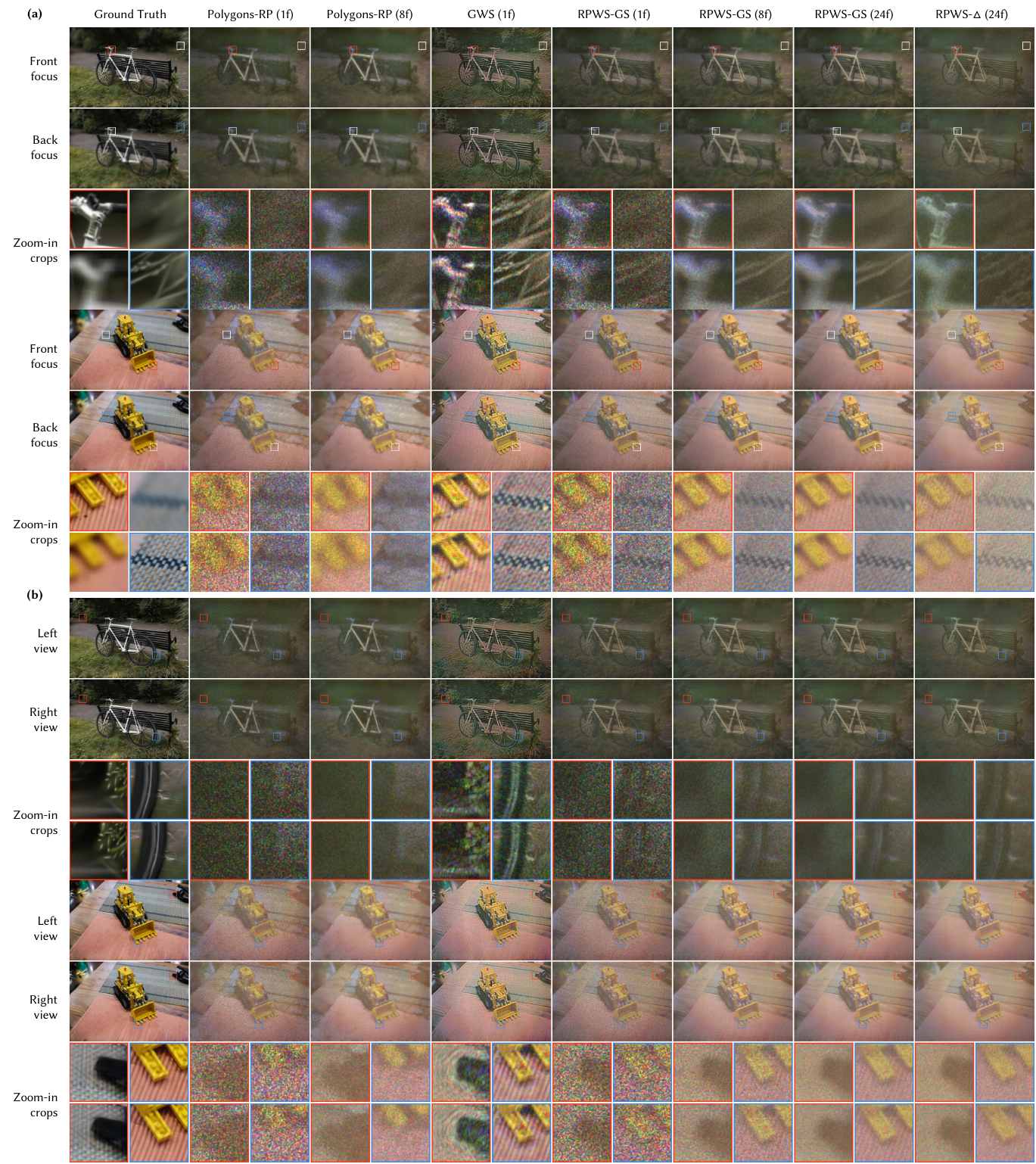}
    \caption{\textbf{Experimentally captured 3D focal stacks and 4D light fields of holograms generated using different CGH algorithms. } Polygon-based CGH (Polygons-RP) \cite{matsushima2005computer, matsushima2009extremely, matsushima2014silhouette} achieves low image quality due to the low quality of the underlying textured mesh 3D representation. GWS \cite{choi2025gaussian} generates smooth-phase holograms, resulting in limited defocus blur with unnatural ringing artifacts and little-to-no parallax. Our method, RPWS with Gaussian (-GS) and triangle splatting ($-\Delta$) variants, achieves good image quality in in-focus regions, reconstructs natural incoherent blur in defocus regions, and shows significantly more parallax than GWS. With 24 frames time-multiplexing, RPWS (both -GS and $-\Delta$ variants) achieves near speckle-free results.}
    \label{fig:fs_captured}
\end{figure*}

We experimentally capture reconstructed focal stacks and \textit{light fields} of RPWS and other CGH baselines to demonstrate real-world refocusing capabilities and parallax effects. Individual frames and color channels are captured independently and merged in post-processing. The experimentally captured results well match the simulation results, as shown in Fig. \ref{fig:fs_captured}. Our method (both RPWS-GS and RPWS-$\Delta$) achieves the best balance between high in-focus image quality and natural defocus blur, while correctly reconstructing parallax.

\label{subsubsec:alpha_blending}
\subsubsection{Ablation Study on Alpha-Blending Operations. } To our knowledge, no effective method was ever presented to alpha-blend \textit{complex, translucent wavefronts}. We thus extend GWS \cite{choi2025gaussian} to support random-phase alpha blending by simply modulating each Gaussian with random phase and running GWS as is, and ablate the wave splatting and intensity-based alpha blending designs in Fig. \ref{fig:alpha_blending_comparison}. Na\"ively adding random phase to GWS (GWS-NRP) fails to reconstruct accurate appearance. With square root alpha blending, GWS yields somewhat plausible results. However, the resulting focal stacks exhibit prominent dark halo and light leakage artifacts at depth discontinuities. In light field reconstruction, this manifests as dark borders around foreground occluders.

Our final RPWS algorithm, which combines the square root alpha-blending formulation \textit{and} the back-to-front wave compositing procedure, achieves the most accurate focal stack reconstruction and parallax rendering results, completely eliminating artifacts at depth discontinuities and occlusion borders while accurately reconstructing image content. Please refer to the supplemental materials for intuitive visual illustrations and explanations of each alpha blending method and why all fail except for ours.

\section{Discussion}
\noindentparagraph{\bf{Limitations and Future Work.}}
In RPWS, time multiplexing is used to realize mathematically exact alpha blending, hence many frames are required for accurate reconstruction. In practice, the effectiveness of time multiplexing is constrained by SLM refresh rates, though this limitation may be alleviated by advances in high-speed SLMs, such as the 3600\,fps FLCoS used in Holographic Parallax~\cite{kim2024holographic} and the recently developed 5760\,fps TI MEMS-based PLM. Addressing quantization artifacts in these devices presents an interesting future direction. Our experiments also show reduced contrast due to random phase, but extending learning-based calibration techniques~\cite{peng2020neural, choi2022time} to directly reconstruct random-phase complex fields~\cite{jang2024waveguide} may further enhance RPWS holograms. Please refer to the supplemental materials for more discussions on contrast enhancement strategies.

\noindentparagraph{\bf{Conclusion.}} Random-phase Wave Splatting enables photorealistic CGH with correct defocus blur and parallax, unlocking perceptually realistic VR. It generalizes beyond Gaussians to diverse translucent primitives, seamlessly bridging modern neural scene representations with next-generation display technology.

\bibliographystyle{ACM-Reference-Format}

\bibliography{bibs}

\end{document}